\documentstyle[prd,aps,floats,epsf,12pt]{revtex}
\tighten

\newcommand{\Fig}[1]{Fig.~(\ref{#1})}
\newcommand{\be}{\begin{equation}}
\newcommand{\ee}{\end{equation}}

\begin{document}
\title{Inhomogeneity and Nonlinear Preheating}

\author{Matthew Parry}

\address{Theoretical Physics Group, Blackett Laboratory, Imperial
College, Prince Consort Rd, London, SW7 2BZ, UK\\E-mail: mparry@ic.ac.uk}

\author{Richard Easther}

\address{Institute for Strings, Cosmology and Astroparticle Physics,
Columbia University, New York, NY 10027,
USA\\Department of Physics, Brown University, Providence, RI 02912,
USA\\ E-mail: easther@physics.columbia.edu}

\maketitle

\vskip0.5cm

\abstract{\bf{We investigated the possibility that nonlinear
gravitational effects influence the preheating era after inflation,
using numerical solutions of the inhomogeneous Einstein field
equations. We compared our results to perturbative calculations and to
solutions of the nonlinear field equations in a rigid (unperturbed)
spacetime, in order to isolate gravitational phenomena.  We confirm
the broad picture of preheating obtained from the nonlinear field
equations in a rigid background, but find gravitational effects have
a measurable impact on the dynamics. The longest modes in the
simulation grow much more rapidly in the relativistic calculation than
with a rigid background. We used the Weyl tensor to quantify the
departure from homogeneity in the universe. We saw no evidence for the
sort of gravitational collapse that leads to the formation of
primordial black holes.}}

Imperial/TP/0-01/9 \qquad Brown-HET-1257 (TA-589) \qquad CU-TP-1002

\vskip0.5cm
For nearly a decade it has been known that coherent processes at
the end of inflation can drive explosive particle production
\cite{TraschenET1990a,KofmanET1994a,ShtanovET1995a} via parametric
resonance, leading to an era of {\em preheating\/} in the
post-inflationary universe. Preheating is intrinsically nonlinear
and takes place far from thermal equilibrium.

A full treatment of parametric resonance must include the
inhomogeneous metric induced by the inhomogeneous matter.
Incorporating metric perturbations into the analysis of parametric
resonance has been the subject of considerable recent attention.
(See \cite{ep99} and references therein.)

As a testbed for preheating, we considered the chaotic
inflationary model driven by the potential $V(\phi) =
\lambda\phi^4/4$ where $\phi$ is the inflaton field. The onset of
resonance can be studied perturbatively, however perturbation
theory cannot give a full description of resonance, since once the
$\delta\phi_k$ grow large, mode-mode couplings become significant.
The next level of sophistication is to include all terms which are
nonlinear in the fields, but to assume that the background FLRW
spacetime is unperturbed. However if the spatial gradients are
non-zero, the metric cannot be perfectly homogeneous, so
gravitational effects on preheating are ignored when this equation
is solved. Therefore one is led finally to numerically solving the
full Einstein field equations.

To reduce the computational complexity of the problem, we assumed
that the universe has a planar symmetry; the metric functions
depend only on $\eta$ and $\zeta$, and are independent of $x$ and
$y$. We used the metric
\be\label{metric2}
ds^2 = \alpha^2(\eta,\zeta)\,(d\eta^2 - d\zeta^2) -
\beta^2(\eta,\zeta)\,(dx^2 + dy^2).
\ee
The equations of motion and details of our numerical code may be
found in \cite{ep99}. We began our simulations at the end of
inflation, so $\phi$ is described by small fluctuations overlaid
on top of the homogeneous zero mode. The fluctuations are quantum
in origin, and are generated by the usual inflationary mechanism.

\begin{figure}[tbp]
\begin{center}
\begin{tabular}{c}
\epsfysize=4cm
\epsfbox{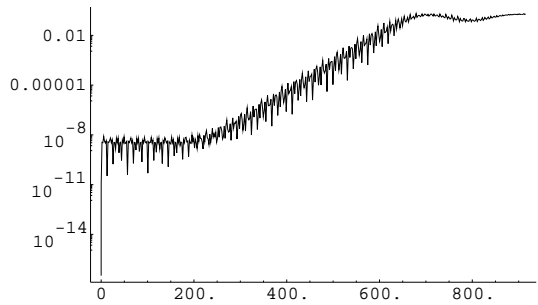}
\end{tabular}
\end{center}
\caption[]{The growth of ${\cal{C}}^2/{\cal{R}}^2$ during the
course of the simulation indicates the growth of inhomogeneity. It
levels off at late times suggesting primordial black holes do not
form. \label{weyl2}}
\end{figure}

For the longest modes there are clear differences between the
results found in a fixed FLRW background and the full relativistic
calculation. Both the relativistic calculation and the nonlinear
field equations in an unperturbed background predict that the
longest modes of the field perturbation grow significantly.
However, the relativistic results predict that this growth begins
earlier and is an order of magnitude larger than that found from
the field evolution in a rigid background. Note that in both cases
we are seeing the growth of modes that lie outside the
instantaneous Hubble horizon.

As a non-perturbative measure of inhomogeneity, we considered the
ratio of ${\cal C}^2$, the spatial average of $C^2=
C_{\mu\nu\gamma\lambda}C^{\mu\nu\gamma\lambda}$, to ${\cal R}^2$,
the spatial average of $R^2$. Because $C^2$ vanishes everywhere in an FLRW
spacetime, the dimensionless quantity ${\cal{C}}^2/{\cal{R}}^2$ is
a measure of how far spacetime departs from a perfectly FLRW
universe. If the enhanced density perturbations generated during
resonance led to localized gravitational collapse and the
formation of primordial black holes (or ``black walls'' in this
case), it would be obvious from ${\cal{C}}^2/{\cal{R}}^2$ as the
curvature would diverge in a collapsing region. Since this does
not happen (see \Fig{weyl2}) we find no evidence that preheating
after $\lambda \phi^4$ inflation leads to primordial black hole
formation.

{\bf Acknowledgments} Computational work in support of this
research was performed at the Theoretical Physics Computing
Facility at Brown University.  RE and MP were supported by DOE
contract DE-FG0291ER40688, Task A.

\end{document}